\documentclass[lettersize,journal]{IEEEtran}
\usepackage{amsmath,amsfonts}
\usepackage{algorithmic}
\usepackage{algorithm}
\usepackage{array}
\usepackage[caption=false,font=normalsize,labelfont=sf,textfont=sf]{subfig}
\usepackage{textcomp}
\usepackage{stfloats}
\usepackage{url}
\usepackage{verbatim}
\usepackage{graphicx}
\usepackage{cite}

\usepackage{xcolor}         

\usepackage{amsmath}
\usepackage{amssymb}
\usepackage{mathtools}
\usepackage{amsthm}

\usepackage{multirow}
\usepackage{comment} 
\usepackage{enumitem}

\usepackage{times}
\usepackage{epsfig}
\usepackage{graphicx}

\usepackage[bb=dsserif]{mathalpha}
\usepackage{bm}

\usepackage{booktabs,colortbl,tabularx}
\usepackage{pifont}%

\definecolor{Gray}{gray}{0.9}
\definecolor{citecolor}{HTML}{2980b9}
\definecolor{linkcolor}{HTML}{c0392b}

\usepackage[pagebackref=true,breaklinks=true,colorlinks,bookmarks=false,citecolor=citecolor,linkcolor=linkcolor]{hyperref}

\usepackage[capitalize,noabbrev]{cleveref}

\newcommand{\cmark}{\ding{51}}%
\newcommand{\xmark}{\ding{55}}%

\begin{document}

\title{Semantic Grouping Network for Audio \\ Source Separation}

\author{Shentong~Mo,~\IEEEmembership{Student Member~IEEE}
        Yapeng~Tian,~\IEEEmembership{Member~IEEE}
\thanks{S. Mo is with Carnegie Mellon University, Pittsburgh, PA, USA.}
\thanks{Y. Tian is with The University of Texas at Dallas, Richardson, TX, USA.}
\thanks{\textit{(Corresponding author: Yapeng Tian, E-mail: yapeng.tian@utdallas.edu.)}}
}

\markboth{Journal of \LaTeX\ Class Files, June~2024}%
{Shell \MakeLowercase{\textit{et al.}}: A Sample Article Using IEEEtran.cls for IEEE Journals}


\maketitle

\begin{abstract}

Recently, audio-visual separation approaches have taken advantage of the natural synchronization between the two modalities to boost audio source separation performance.
They extracted high-level semantics from visual inputs as the guidance to help disentangle sound representation for individual sources.
Can we directly learn to disentangle the individual semantics from the sound itself?
The dilemma is that multiple sound sources are mixed together in the original space. 
To tackle the difficulty, in this paper, we present a novel Semantic Grouping Network, termed as SGN, that can directly disentangle sound representations and extract high-level semantic information for each source from input audio mixture.
Specifically, SGN aggregates category-wise source features through learnable class tokens of sounds. Then, the aggregated semantic features can be used as the guidance to separate the corresponding audio sources from the mixture. 
We conducted extensive experiments on music-only and universal sound separation benchmarks: MUSIC, FUSS, MUSDB18, and VGG-Sound. The results demonstrate that our SGN significantly outperforms previous audio-only methods and audio-visual models without utilizing additional visual cues.

\end{abstract}

\begin{IEEEkeywords}
audio-visual source separation, sound source separation
\end{IEEEkeywords}

\section{Introduction}

\IEEEPARstart{I}{n} our daily life, sound often appears to be a mixture of multiple sources, such as the sound in a classroom scene that consists of professors' speech and students whispering; the audio in a family reunion includes talking, laughing, and background music.
Humans are capable of separating individual sources from these complex mixtures. For example, when we enjoy a beautiful melody, we are naturally aware of how many instruments are playing.
This crucial human perception intelligence of auditory scenes attracts many researchers to explore the audio source separation problem.

Early audio-only works leveraged traditional machine learning approaches, such as  hidden Markov models~\cite{roweis2000one,mysore2010non}, Non-negative Matrix Factorization~\cite{Smaragdis2003nonnegative,Virtanen2007monaural,Cichocki2009nonnegative} and Robust Principal Component Analysis~\cite{huang2012rpca}, to solve the separation task. 
Benefiting from advances in deep learning, deep neural architectures, such as U-Net~\cite{Ulyanov2018unet} have been widely exploited to boost audio separation performance.
Wave-U-Net~\cite{stoller2018waveunet} adopted U-Net to the one-dimensional time domain to resample feature maps at multiple time scales for music source separation.
In order to improve phase reconstruction, ResUNetDecouple+~\cite{Kong2021DecouplingMA} decoupled complex ideal ratio mask estimation into magnitude and phase estimation and proposed a residual U-Net architecture.
Recently, some works~\cite{wisdom2020unsupervised,wisdom2021fuss} have started to explore source separation on universal audio mixtures in open domains. 
However, current audio-only methods cannot learn compact representations for individual sources if just learning a direct mapping from a mixture to individual sources.
In contrast, we will solve them in our approach by extracting disentangled and compact representations as guidance for separation.

Since audio and visual contents are commonly matched and synchronized, audio-visual source separation methods~\cite{hershey2001audio,zhao2018the,ephrat2018looking,Gao2018learning,xu2019mpnet,Gan2020music,tian2021cyclic} are developed to improve separation.
Unlike in audio space, sound sources are naturally separated in visual space. 
Thus, these methods can extract discriminative semantic information from visual inputs to help disentangle sound representation for reconstructing individual audio sources.
SoP~\cite{zhao2018the} extracted pixel-level visual features from a dilated ResNet~\cite{he2016resnet} to select the spectral components associated with the pixels.
They then combined the magnitude of spectrogram with the phase of input spectrogram and applied inverse Short-Time Fourier Transform (STFT) to reconstruct the source waveform.
A MinusPlus Net~\cite{xu2019mpnet} was utilized to subtract salient sounds from the mixture and separate sounds recursively.
A cyclic co-learning framework~\cite{tian2021cyclic} was proposed to separate visual sounds sources with the help of sounding object visual grounding.
While the methods achieve promising results on audio-visual source separation, they are extremely dependent on the ability of visual networks with large parameters. 
Without visual clues, their performance degrades significantly as observed in our experiments.

The success in audio-visual sound separation inspires us to ask a question: \textit{can we directly disentangle the individual semantics from sound itself to guide separation}?
The main challenge is that sounds are naturally mixed in the audio space. The individual semantics can be extracted from separated clean sources, but separation is the goal of our task. It is like a chicken or the egg problem. 
To tackle the dilemma, our key idea is to disentangle individual source representation using semantic-aware grouping to guide separation, which is different from the existing audio-only and audio-visual methods. 
During training, we aim to learn categorical codes to help disentangle and aggregate category-wise source features from the mixture for separation. The features will carry separated high-level semantic information for individual sources.

To this end, we propose a novel Semantic Grouping Network (SGN) that can learn to disentangle individual semantics from sound itself to guide source separation from audio mixtures.
Specifically, we first learn class-aware features through category-aware grouping in terms of learnable source class tokens. 
Meanwhile, we use a U-Net to extract audio embeddeings from the input mixture. Our category-wise representations will serve as the semantic guidance to select their semantically corresponded audio features to reconstruct the corresponding audio sources. Our new framework is simple and flexible. 
During inference, it can separate different number of sources with respect to the aggregated category-specific semantics rather than a fixed number in existing audio-only separation methods.
In our implementation, audio waveforms are converted to spectrograms using Short-time Fourier transform (STFT) for model learning and the final waveforms of individual audio sources can be reconstructed by inverse STFT.

Experimental results on both MUSIC~\cite{zhao2018the} and FUSS~\cite{wisdom2021fuss}, MUSDB18~\cite{musdb18}, and VGG-Sound~\cite{chen2020vggsound} benchmarks can validate the superiority of our SGN against state-of-the-art audio source separation methods.
Notably, without additional visual clues, our audio-only models can even achieve significant gains over audio-visual sound source separation methods.
In addition, qualitative visualizations of separation results vividly showcase the effectiveness of our SGN in separating audio sources from mixtures.
Extensive ablation studies also demonstrate the importance of category-aware grouping and learnable class tokens on learning compact representations for audio source separation.

Our main contributions can be summarized as follows:

\begin{itemize}
    \item We present a novel Semantic Grouping Network, namely SGN, to disentangle the individual semantics from sound itself to guide source separation.
    \item We introduce learnable source class tokens in audio separation to aggregate category-wise source features with explicit high-level semantics.
    \item Extensive experiments on both music-only and universal separation datasets demonstrate the superiority of our SGN over state-of-the-art separation approaches.
\end{itemize}

\section{Related Work}

\noindent\textbf{Audio Representation Learning.}
Audio representation learning has been addressed in many previous works~\cite{aytar2016soundnet,owens2016ambient,Arandjelovic2017look,korbar2018cooperative,Senocak2018learning,zhao2018the,zhao2019the,Gan2020music,Morgado2020learning,Morgado2021robust,Morgado2021audio,mo2023diffava,mo2023oneavm,mo2023deepavfusion,mo2024texttoaudio,mo2023class,pian2023audiovisual} to learn discriminative representations of waveform and spectrogram from audios.
Such learnable features are beneficial for many audio-relevant tasks, such as audio-event localization~\cite{tian2018ave,lin2019dual,wu2019dual,lin2020audiovisual}, audio-visual parsing~\cite{tian2020avvp,wu2021explore,lin2021exploring,mo2022multimodal,mo2022semantic}, audio-visual spatialization~\cite{Morgado2018selfsupervised,gao20192.5D,Chen2020SoundSpacesAN,Morgado2020learning}, and sound source localization~\cite{Senocak2018learning,hu2019deep,Afouras2020selfsupervised,qian2020multiple,chen2021localizing,mo2022EZVSL,mo2022slavc,mo2022benchmarking,mo2023avsam,mo2023audiovisual,mo2023weaklysupervised}.
In this work, our main focus is to learn compact audio representations for source separation from sound mixtures, which is more challenging than those tasks listed above.

\noindent\textbf{Audio-Visual Source Separation.}
Audio-visual source separation aims at separating sound sources from the audio mixture given the image of sources, such as music source separation from a picture of orchestra playing.
In the recent years, researchers~\cite{hershey2001audio,zhao2018the,ephrat2018looking,Gao2018learning,xu2019mpnet,tian2021cyclic,tzinis2020into} have proposed diverse pipelines to learn better visual representations from images.
Zhao \textit{et al.}~\cite{zhao2018the} first leveraged visual features to learn the correspondance between the spectral components and pixels for recovering magnitude and phase of input spectrogram.
MP-Net~\cite{xu2019mpnet} applied a recursive MinusPlus Net to separate salient sounds from the mixture.
Tian \textit{et al.}~\cite{tian2021cyclic} leveraged sounding object visual grounding to separate visual sound sources in a cyclic co-learning framework.
To make full use of visual clues, more modalities are introduced to learn better visual represenations, such as motion in SoM~\cite{zhao2019the}, gesture consisting of pose and keypoints in MG~\cite{Gan2020mg}, and spatio-temporal visual scene graphs in AVSGS~\cite{Chatterjee2021visual}.
While the aforementioned audio-visual approaches achieve promising performance on source separation, they are extremely dependent on the ability of visual networks with exhaustive parameters to learn discriminative representations. 
Without visual features as clues, their performance degrades significantly as observed in
our experiments in Sec.~\ref{sec:exp}.
In this work, we aim to get rid of the dependency of visual sounds separation on visual branches, by introducing learnable source class tokens to learn compact audio representations for separation.

\noindent\textbf{Audio Source Separation.}
Audio source separation is a challenging problem that separates the individual audio source from a mixture without any visual clues.
Early methods applied classical hidden Markov models~\cite{roweis2000one,mysore2010non}, Non-negative Matrix Factorization~\cite{Smaragdis2003nonnegative,Virtanen2007monaural,Cichocki2009nonnegative}, Robust Principal Component Analysis~\cite{huang2012rpca}, and clustering~\cite{Hershey2016deep} to separate the source component directly from the mixture.
With the development of deep neural networks,  U-Net~\cite{Ulyanov2018unet} with learnable parameters was introduced to learn the spectrogram representations of audio
mixtures.
Wave U-Net~\cite{stoller2018waveunet} proposed to learn multi-scale feature maps from the one-dimensional time domain of waveform.
Recently, ResUNetDecouple+~\cite{Kong2021DecouplingMA} used a residual U-Net to decouple the estimation of complex ideal ratio masks into magnitude and phase estimations for music sources.
Beyond music source separation, MixIT~\cite{wisdom2020unsupervised} optimizes to map estimated sources to reference mixtures for separation in the large-scale YFCC100M~\cite{thomee2016yfcc100m} videos.
More recently, a task of single channel distance-based sound separation~\cite{patterson2022distance} was proposed to separate near sounds from far sounds in synthetic reverberant mixtures.
Different from audio source separation baselines, we develop a fully novel framework to aggregate compact category-wise audio source representations with explicit learnable source class tokens.
To the best of our knowledge, we are the first to leverage an explicit grouping mechanism for audio source separation.
In addition, we do not need the unsupervised learning on large-scale simulated audio data with expensive training costs.
Our experiments in Sec.~\ref{sec:exp} also demonstrate the effectiveness of SGN in source separation on both music-only and universal sounds.

\noindent\textbf{Weakly-Supervised Audio Learning.}
Weak supervision has been proven successful in diverse audio-relevant tasks, such as sound events detection~\cite{Anurag2016AudioEA,Kumar2016audio,Shah2018ACL} and sound source separation~\cite{Seetharaman2019class,Pishdadian2019FindingSI,Tzinis2020improving}.
Typically, WAL-Net~\cite{Shah2018ACL} proposed using weakly labeled data
from the web to explore how the label density and corruption of labels affect the generalization of models.
DC/GMM~\cite{Seetharaman2019class} applied an auxiliary network to generate the parameters of Gaussians in the embedding space with a one-hot vector indicating the class as input.
However, these approaches just learn a direct mapping from a mixture to individual sources, which cannot capture compact representations
for individual sources.
In contrast, we will solve them in our approach by leveraging learnable source class tokens to extract disentangled and compact representations as guidance for separation.

\begin{figure*}[t]
\centering
\centering
    \includegraphics[width=0.7\linewidth]{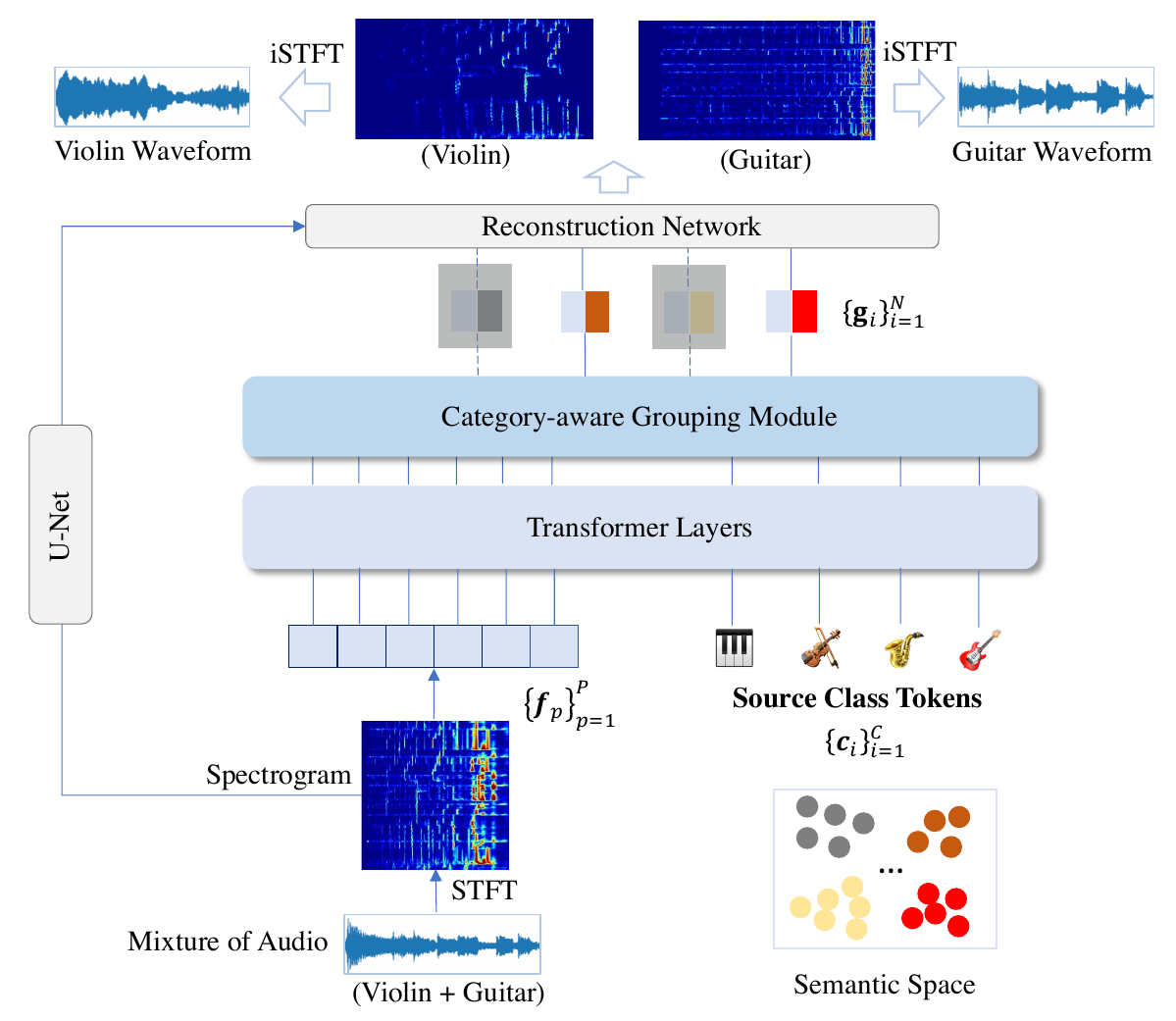}
\caption{Illustration of our Semantic Grouping Network (SGN).
The Category-aware Grouping module takes as input raw features of the input mixture spectrogram $\{\mathbf{f}_p\}_{p=1}^P$ and learnable class tokens $\{\mathbf{c}_i\}_{i=1}^C$ of for $C$ categories in the semantic space to generate disentangled class-aware representations $\{\mathbf{g}_i\}_{i=1}^N$ for $N$ sources.
Note that $N$ source embeddings are chosen according to the ground-truth class.
With the category-aware embeddings $\{\mathbf{g}_i\}_{i=1}^N$ and U-Net features of the mixture spectrogram, a
light reconstruction network is used to reconstruct source spectrograms.
Finally, inverse STFT is applied to recover the waveform of each source from spectrograms.
}
\label{fig: main_img}
\end{figure*}

\section{Method}

Given a mixture of audios, our target is to separate all individual sound sources. 
We propose a novel Semantic Grouping Network named SGN for disentangling individual semantics from sound itself, which mainly consists of two modules, Source Class Tokens in Section~\ref{sec:sct} and Category-aware Grouping in Section~\ref{sec:cag}.

\subsection{Preliminaries}

In this section, we first describe the problem setup and notations, and then revisit an audio-visual separation method: Sound of Pixels~\cite{zhao2018the}.

\noindent\textbf{Problem Setup and Notations.}
Given a mixture of audio waveform, we apply Short-Time Fourier Transform (STFT) to extract a sound spectrogram with time $T$ and frequency $F$ from it.
Our goal is to recover a matrix $\mathbf{W}$ of sound spectrograms with $N$ sources, where $\mathbf{W} \in \mathbb{R}^{TF\times N}$.
Note that $T, F$ denotes the dimension of time and frequency of each source spectrogram.

\noindent\textbf{Revisit Sound of Pixels.}
To solve the audio source separation problem, SoP~\cite{zhao2018the} extracted pixel-level visual features $\mathbf{P}$ to select the spectral components corresponding to the pixel, by leveraging the audio-visual synchronization.
Since sounding sources are naturally separated in visual space, applying pixel-level features $\mathbf{P}$ on U-Net~\cite{Ulyanov2018unet} features $\mathbf{U}$ of the mixture spectrogram to recover the source matrix $\mathbf{W}$, which is denoted as
\begin{equation}\label{eq:sop}
    \mathbf{W} = \mathbf{U}\mathbf{P}
\end{equation}
where $\mathbf{U}\in\mathbb{R}^{TF\times D},\mathbf{P}\in\mathbb{R}^{D\times N}$, and $D$ is the dimension size of features.
With the benefit of disentangled visual features, audio-visual separation methods achieved promising results on sound source separation.

However, those audio-visual approaches are extremely dependent on the capacity of the pre-trained visual networks to extract disentangled features. 
Without visual cues, their performance deteriorates significantly as shown in Sec~\ref{sec:exp}.
To address this issue, we are motivated by~\cite{xu2022groupvit} and propose a novel Semantic Grouping Network that can learn to disentangle the individual semantics from sound itself to guide source separation from audio mixtures, as illustrated in Figure~\ref{fig: main_img}.

\subsection{Source Class Tokens}\label{sec:sct}

In order to explicitly disentangle individual semantics from the mixed sound space, we introduce learnable source-specific class tokens $\{\mathbf{c}_i\}_{i=1}^C$ to help group semantic-aware information from raw features $\{\mathbf{f}_{p}\}_{p=1}^P$ that are extracted from the input mixture spectrogram via a convolution patch embedding layer~\cite{dosovitskiy2020image}, where $\mathbf{c}_i\in\mathbb{R}^{1\times D}, \mathbf{f}_{p}\in\mathbb{R}^{1\times D}$, $C$ is the total number of audio source classes, $P$ is the total number of patches, and $D$ is the dimension size.

With the categorical token embeddings, we first apply self-attention transformers $\phi(\cdot)$ to aggregate temporal features from the raw input as
\begin{equation}
\begin{aligned}
    \{\hat{\mathbf{f}}_{p}\}_{p=1}^P, \{\hat{\mathbf{c}}_i\}_{i=1}^C = \{\phi(\mathbf{x}_j, \mathbf{X}, \mathbf{X})\}_{j=1}^{P+C}, \\
    \mathbf{X} = \{\mathbf{x}_j\}_{j=1}^{P+C} = [\{\mathbf{f}_{p}\}_{p=1}^P; \{\mathbf{c}_i\}_{i=1}^C]
\end{aligned}
\end{equation}
where $[\ ;\ ]$ denotes the concatenation operator.
$\mathbf{f}_p,\mathbf{c}_i,\mathbf{x}_j\in\mathbb{R}^{1\times D}$, and $D$ is the dimension of embeddings. 
The self-attention operator $\phi(\cdot)$ is formulated as:
\begin{equation}
    \phi(\mathbf{x}_j, \mathbf{X}, \mathbf{X}) = \mbox{Softmax}(\dfrac{\mathbf{x}_j\mathbf{X}^\top}{\sqrt{D}})\mathbf{X}
\end{equation}
Then, to constrain the independence of each class token $\mathbf{c}_i$ in the semantic space, we exploit a fully-connected (FC) layer and add softmax operation to predict the class probability: $\mathbf{e}_i = \mbox{Softmax}(\textsc{FC}(\mathbf{c}_i))$. 
Each category probability is constrained by a cross-entropy loss
$\sum_{i=1}^C\mbox{CE}(\mathbf{h}_i, \mathbf{e}_i)$, where $\mbox{CE}(\cdot)$ is cross-entropy loss; $\mathbf{h}_i$ denotes an one-hot encoding and only its element for the target category entry $i$ is 1. With optimizing the loss, it will push the learned token embeddings to be discriminative and category-specific.

\subsection{Category-aware Grouping}\label{sec:cag}

Benefiting from the class-constraint loss mentioned above, we propose a novel and explicit category-aware grouping module $g(\cdot)$ to take audio mixture features and class tokens as inputs to generate source category-aware representations as:
\begin{equation}
    \{\mathbf{g}_i\}_{i=1}^C = \{g(\{\hat{\mathbf{f}}_{p}\}_{p=1}^P, \hat{\mathbf{c}}_i)\}_{i=1}^C
\end{equation}
In the time of grouping, we conflate features from the same source category token into the new category-aware representation according to a similarity matrix $\mathbf{A}$ between features and category token, which is calculated as:
\begin{equation}
    \mathbf{A}_{p,i} = \mbox{Softmax}(W_q\hat{\mathbf{f}}_p, W_k\hat{\mathbf{c}}_i)
\end{equation}
where $W_q\in\mathbb{R}^{D\times D}, W_k\in\mathbb{R}^{D\times D}$ are learnable weights of linear projection layers for features and category tokens.
With the similarity matrix, we compute the weighted sum of temporal features to generate the category-aware semantic representation for category $i$ as:
\begin{equation}
    \mathbf{g}_i = g(\{\hat{\mathbf{f}}_{p}\}_{p=1}^P, \hat{\mathbf{c}}_i) = \hat{\mathbf{c}}_i + W_o\dfrac{\sum_{p=1}^P\mathbf{A}_{p,i}W_v\hat{\mathbf{f}}_{p}}{\sum_{p=1}^P\mathbf{A}_{p,i}}
\end{equation}
where $W_o\in\mathbb{R}^{D\times D}, W_v\in\mathbb{R}^{D\times D}$ are learnable weights of linear projection layers for output and value.
Taken category-aware representations $\{\mathbf{g}_i\}_{i=1}^C$ as the inputs, we use a FC layer and sigmoid operation to predict the binary probability: $p_i = \mbox{Sigmoid}(\textsc{FC}(\mathbf{g}_i))$ for $i$th category. 
By applying audio source classes $\{y_i\}_{i=1}^C$ as the weak supervision and combining the class-constraint loss, we formulate a semantic-aware grouping loss as:
\begin{equation}
    \mathcal{L}_{\mbox{group}} = \sum_{i=1}^C\{\mbox{CE}(\mathbf{h}_i, \mathbf{e}_i) + \mbox{BCE}(y_i, p_i)\}.
\end{equation}
Since multiple audio sources could be in one mixture, we use binary cross-entropy loss: $\mbox{BCE}(\cdot)$ for each cateogry to handle this multi-label classification problem.

Similar to SoP~\cite{zhao2018the}, we adopt a binary mask matrix $\mathbf{M}\in\mathbb{R}^{TF\times N}$ that separates the mixture spectrogram as the final target for stabilized training, that is, $\mathbf{M}=\{\mathbf{m}_i\}_{i=1}^N, \mathbf{m}_i\in\mathbb{R}^{TF}$.
The binary mask $\mathbf{m}_i$ is computed by predicting whether the $i$th source is the dominant component in the mixed sound as $\mathbf{m}_i = \mathbf{w}_i>0.5\mathbf{w}_{mix}$, where $\mathbf{w}_{i}, \mathbf{w}_{mix}\in\mathbb{R}^{TF}$ denotes the $i$th source and mixture spectrogram, respectively.
With the category-aware embeddings $\{\mathbf{g}_i\}_{i=1}^C$ and U-Net features $\mathbf{U}$ of the mixture spectrogram as inputs, we adopt a light reconstruction network composed of an inner product operator and learnable bias parameters to reconstruct spectrogram masks $\{\hat{\mathbf{m}}_i\}_{i=1}^N$ with $N$ sources.
Finally, a reconstruction loss is formulated with the sum of binary cross-entropy for $N$ sources as:
\begin{equation}
    \mathcal{L}_{\mbox{rec}} = \sum_{i=1}^N\mbox{BCE}(\mathbf{m}_i, \hat{\mathbf{m}}_i)
\end{equation}
where $N$ denotes the number of mixture and $m_i$ refers to the ground-truth mask. 
The overall objective of our model is simply optimized in an end-to-end manner as:
\begin{equation}
    \mathcal{L} = \mathcal{L}_{\mbox{rec}} + \mathcal{L}_{\mbox{group}} 
\end{equation}
During inference, we multiply the predicted $N$ individual masks: $\{\hat{\mathbf{m}}_i\}_{i=1}^N\in\mathbb{R}^{TF\times N}$ by the mixture spectrogram $\mathbf{W}_{mix}\in\mathbb{R}^{TF}$ to generate the audio spectrograms for separated sources, that is, $ \{\hat{\mathbf{w}}_i\}_{i=1}^N=\{\mathbf{W}_{mix}\odot\hat{\mathbf{m}}_i\}_{i=1}^N$, where $\odot$ denotes element-wise multiplication. $\hat{\mathbf{w}}_i$ is reshaped to a $T\times F$ spectrogram and inverse STFT is applied to recover the waveform of each audio source.

\noindent\textbf{Relation to Sound of Pixels.}
Recall that SoP~\cite{zhao2018the} separated U-Net features $\mathbf{U}\in\mathbb{R}^{TF\times D}$ into source matrix $\mathbf{W}$ and visual features $\mathbf{P}$ in Eq.~\ref{eq:sop}.
In contrast to their solutions, the main motivation of our SGN is to separate U-Net features $\mathbf{U}$ into source matrix $\mathbf{W}$ with respect to disentangled category-aware representations $\mathbf{G}$, which is formulated as:
\begin{equation}
    \mathbf{W} = \mathbf{U}\mathbf{G}
\end{equation}
where $\mathbf{G}\in\mathbb{R}^{D\times N}$ with $N$ source class-specific embeddings.
Different from SoP~\cite{zhao2018the} using visual cues, the proposed SGN enables learning disentangled semantics from audio itself to guide sound source separation from mixtures.

\begin{table*}[!tb]
    \renewcommand{\arraystretch}{1.1}
	\centering
	\caption{Quantitative results of audio source separation on MUSIC dataset for two sources.}
	\label{tab: exp_music}
	\scalebox{0.98}{
		\begin{tabular}{lcccc}
			\toprule
			Method & Input & SDR & SIR & SAR \\
			\midrule
			RPCA~\cite{huang2012rpca} & audio-only & -0.62 &	2.32 &	2.41 \\
			NMF~\cite{Virtanen2007monaural} & audio-only & 0.86 &	3.26 &	3.81 \\
			SoP~\cite{zhao2018the} (w/o visual) & audio-only & 2.16 &	5.58 &	5.54 \\
			Wave-U-Net~\cite{stoller2018waveunet} & audio-only & 3.80 &	6.75 &	6.62 \\
			ResUNetDecouple+~\cite{Kong2021DecouplingMA} & audio-only & 3.98 &	7.17 &	6.91 \\ \hline
			SoP~\cite{zhao2018the} & audio-visual & 4.55 & 10.06 & 10.24 \\
			MP-Net~\cite{xu2019mpnet} & audio-visual & 4.82 & 10.19 & 10.56 \\ \hline
                Wave-U-Net~\cite{stoller2018waveunet} + class & audio-only & 4.02	& 7.51 & 7.05 \\
                ResUNetDecouple+~\cite{Kong2021DecouplingMA} + class & audio-only &  4.53 & 8.62 & 7.82 \\ \hline
			SGN (ours) & audio-only & \textbf{5.20} & \textbf{10.81} & \textbf{10.67} \\
			
			\bottomrule
			\end{tabular}}
\end{table*}

\section{Experiments}

\subsection{Experimental Setup}

\noindent\textbf{Datasets.}
MUSIC\footnote{Since many videos are no longer publicly available, the used dataset is smaller than the original MUSIC dataset. For a fair comparison, we trained all models on the same training data.}~\cite{zhao2018the} contains 448 untrimmed YouTube music videos of solos and duets from 11 instrument categories.
358 solo videos are applied for training, and 90 solo videos for evaluation.
FUSS~\cite{wisdom2021fuss} is a universal sound dataset with 10 second clips from FSD50K~\cite{Fonseca2017freesound} with annotated labels from the AudioSet Ontology, which includes between 1 and 4 sound sources.
The number of available categories is 286.
We use 20000 mixture clips for training, 1000 mixture clips for validation, and  1000 mixture clips for testing.
MUSDB18~\cite{musdb18} includes 50 full lengths of music tracks with 10h duration of different genres along with isolated drums, bass, vocals, and other stems.
Beyond those datasets, we filter 150k video clips of 10s lengths from the original VGG-Sound~\cite{chen2020vggsound}, which contains 221 categories, such as animals, instruments, vehicles, people, etc.

\noindent\textbf{Evaluation Metrics.}
Following previous work~\cite{zhao2018the,roux2018sdr,stoller2018waveunet,Kong2021DecouplingMA}, we use scale-invariant SDR (SI-SDR), Signal-to-Distortion Ratio (SDR), Signal-to-Interference Ratio (SIR), and Signal-to-Artifact Ratio (SAR) to evaluate the separation performance.
The open-source mir\_eval~\cite{raffel2014mireval} library is utilized for computing the results to report.

\noindent\textbf{Network Structures.}
In this part, we report detailed network structures and parameters used in the proposed SGN.
First, we use a Convolution2D layer to extract raw features from the input mixture spectrogram with a shape of $256\times 256$, where the kernel size is $16$ and stride size is $16$. The dimension size of embeddings is $256$.
Then we adopt 6 self-attention transformer layers to extract audio representations $256\times 256$ and class embeddings $11\times 256$.
In Category-aware Grouping module, we utilize class embeddings to aggregate new category-aware representations $11\times 256$.
Meanwhile, the U-Net architecture with $7$ downsampling blocks and $7$ upsampling blocks in SoP~\cite{zhao2018the} is applied to extract audio mixture features $65536\times 256$.
In the end, we leverage a light reconstruction network composed of inner product on new category-aware embeddings and audio mixture features to generate category-aware spectrograms $11\times 65536$.
For each source reconstruction, we keep the corresponding class index from $11\times 65536$ to use the output $65536$ as the final spectrogram (reshaped as $256\times 256$).

\noindent\textbf{Implementation.}
We follow the prior work and the audio signal is sub-sampled to 11kHz.
An STFT with a window size of 1022 and a hop length of 256 is further applied on to generate a 512 × 256 Time-Frequency representation of the audio, which is resampled to a log-frequency scale with size of 256 × 256 as the input spectrogram.
A SGD optimizer with momentum 0.9 is used to train the model.
The learning rate is 0.001.
We train the model with a batch size of 80 for 100 epochs.
The depth of self-attention transformers is 6, and the dimension size $D$ is 256.
The kernel and stride size of the convolution patch embedding layer is 16 and 16.
The total number of patches $P$ is 256.
In experiments, the number of sources $N$ for training is 2.

\begin{table*}[!tb]
    \renewcommand{\arraystretch}{1.1}
	\centering
	\caption{Quantitative results of audio-only universal sound separation on FUSS dataset for two sources.}
	\label{tab: exp_fuss}
	\scalebox{0.98}{
		\begin{tabular}{lccccc}
			\toprule
			Method & Input & SI-SDR & SDR & SIR & SAR \\
			\midrule
			RPCA~\cite{huang2012rpca} & audio-only & -0.52  &   -1.16 & 0.28 & 1.24 \\
			NMF~\cite{Virtanen2007monaural} & audio-only &  0.32 & -0.49 & 0.56 & 2.80 \\
			SoP~\cite{zhao2018the} (w/o visual) & audio-only & 2.51   & 1.65 & 2.98 & 3.25 \\ 
			Wave-U-Net~\cite{stoller2018waveunet} & audio-only & 3.52 & 2.36 & 5.12 & 4.89 \\
			ResUNetDecouple+~\cite{Kong2021DecouplingMA} & audio-only &  3.76 &  2.57 & 5.63 & 5.38 \\
			TDCN++~\cite{wisdom2021fuss} & audio-only & 4.51 & 3.93 & 7.21 & 7.66 \\
			\hline
                Wave-U-Net~\cite{stoller2018waveunet} + class & audio-only & 3.87 & 2.63 & 5.78 & 5.57 \\
                ResUNetDecouple+~\cite{Kong2021DecouplingMA} + class & audio-only & 4.32 & 3.26 & 6.51 & 6.22 \\ \hline
			SGN (ours) & audio-only & \bf 5.67 & \textbf{4.56} & \textbf{8.02} & \textbf{8.47} \\
			
			\bottomrule
			\end{tabular}}
\end{table*}

\subsection{Comparison to Prior Work}\label{sec:exp}

In this work, we propose a novel and effective framework for audio source separation. 
In order to validate the effectiveness of the proposed SGN, we fully compare it to previous audio-only and audio-visual baselines:
1) NMF~\cite{Virtanen2007monaural}: a traditional signal processing baseline with  non-negative matrix factorization to separate each testing source spectrogram directly;
2) RPCA~\cite{huang2012rpca}: a machine learning method without parameters to separate each source via robust principal component analysis;
3) SoP (w/o visual)~\cite{zhao2018the}: an audio-visual baseline without the visual network to do separation by applying U-Net on audio only;
4) WAVE U-NET~\cite{stoller2018waveunet}: an audio-only approach with multi-scale feature maps from one-dimensional waveform;
5) SoP~\cite{zhao2018the}: the first audio-visual model on MUSIC dataset by leveraging pixel-level visual features to match the spectral components to recover magnitude and phase of input spectrogram;
6) MP-Net~\cite{xu2019mpnet}: an audio-visual baseline with recursive separation from the mixture;
7) ResUNetDecouple+~\cite{Kong2021DecouplingMA}: a recent method by applying a residual U-Net on spectrogram to decouple the estimation of complex ideal ratio mask into phase and magnitude;
8) TDCN++~\cite{wisdom2021fuss}: a typical audio-only baseline for universal sound source separation on FUSS benchmark.
As previous category-based methods~\cite{Seetharaman2019class,Pishdadian2019FindingSI,Tzinis2020improving} are not open-sourced, we add a classifier with category as supervision to WAVE U-NET~\cite{stoller2018waveunet} and ResUNetDecouple+~\cite{Kong2021DecouplingMA} for a comprehensive comparison.

For music-only sound source, we report the quantitative comparisons in Table~\ref{tab: exp_music}.
As can be seen, we achieve the best performance in terms of all metrics compared to previous both audio-only and audio-visual baselines.
In particular, the proposed SGN significantly outperforms ResUNetDecouple+~\cite{Kong2021DecouplingMA}, the current state-of-the-art audio-only model, where we achieve the performance gains of 1.22 SDR, 3.64 SIR, and 3.76 SAR.
Without addition visual cues, the performance of SoP~\cite{zhao2018the} drops a lot, which implies the importance of extracting high-level semantics from visual inputs as the guidance for audio-visual approaches.
Meanwhile, our SGN achieves comparable even better results against those audio-visual baselines and category-based methods.
These improvements demonstrate the effectiveness of our method in disentangling individual semantics from the audio itself.

In addition, significant gains on universal audio source separation can be observed in Table~\ref{tab: exp_fuss}.
Compared to ResUNetDecouple+~\cite{Kong2021DecouplingMA} with decoupling estimations of phase and magnitude, our SGN achieves the results gains of 1.99 SDR, 2.39 SIR, and 3.09 SAR.
Furthermore, the proposed approach outperforms the TDCN++ baseline~\cite{wisdom2021fuss} by 0.63 SDR, 0.81 SIR, and 0.81 SAR. 
When evaluating comparisons using SI-SDR, we achieve the results of 5.67, which significantly outperforms the state-of-the-art audio-only approach (ResUNetDecouple+~\cite{Kong2021DecouplingMA}) with classes as weak supervision (4.32 SI-SDR).
We also achieve better performance than category-based approaches in terms of all metrics.
These comparison results validate the superiority of our method in sound separation.

We also conduct experiments on the MUSDB18~\cite{musdb18} benchmark and achieved an SI-SDR of 10.23, while the state-of-the-art audio-only approach (ResUNetDecouple+) achieved 8.15 SI-SDR with the ground-truth classes as supervision.
We further evaluate the proposed method on the original VGG-Sound benchmark~\cite{chen2020vggsound}, including 221 categories, such as animals, instruments, vehicles, people, etc.
Under the 2-source mixture setting, our SGN (4.85 SDR, 8.83 SIR, 8.92 SAR) significantly outperforms ResUNetDecouple+~\cite{Kong2021DecouplingMA}(2.43 SDR, 5.01 SID, 4.86 SAR), the current state-of-the-art audio-only model, where we achieve the performance gains of 2.42 SDR, 3.82 SIR, and 4.06 SAR.

\begin{figure*}[t]
\centering
\includegraphics[width=0.97\linewidth]{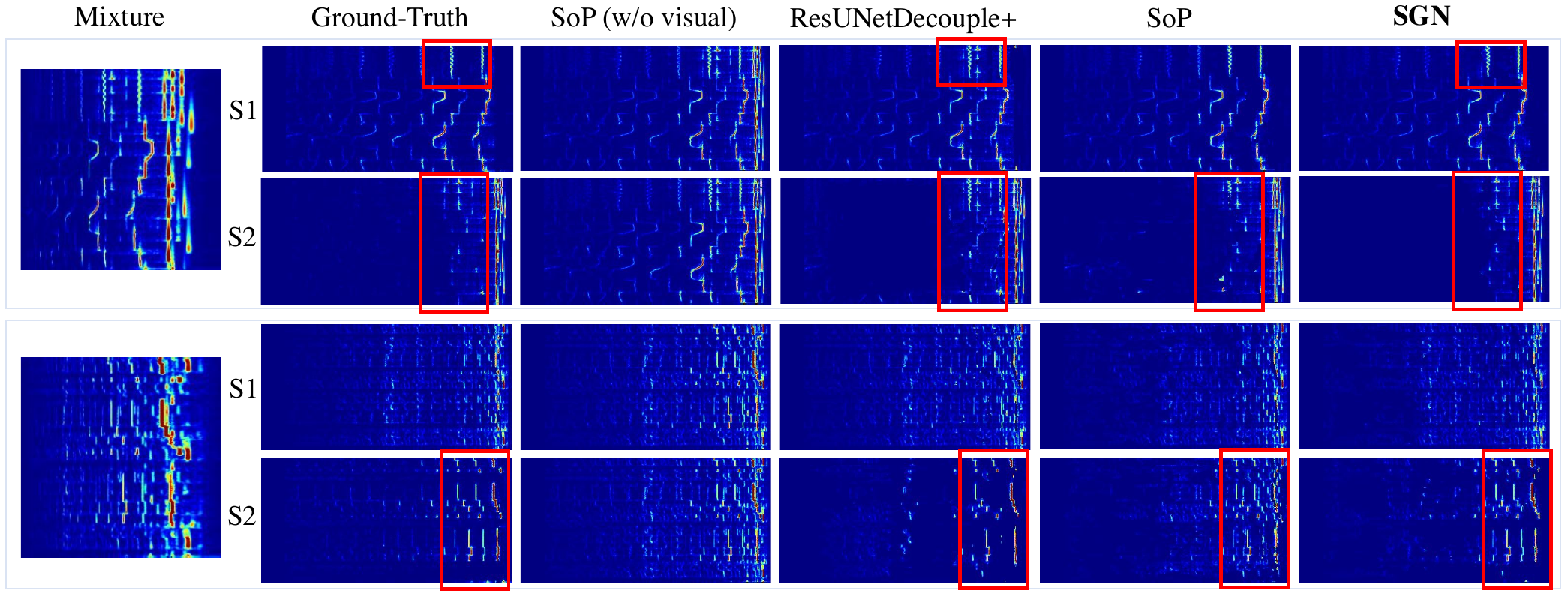}
\caption{Qualitative comparisons with audio-visual and audio-only baselines~\cite{zhao2018the,Kong2021DecouplingMA}. The proposed SGN achieves much better separation performance in terms of the quality of reconstructed source spectrograms. To better illustrate the superiority, some regions are highlighted by red boxes.
}
\label{fig: exp_vis}
\end{figure*}

In order to qualitatively evaluate reconstructed source spectrograms, we compare the proposed SGN with audio-visual and audio-only baselines~\cite{zhao2018the,Kong2021DecouplingMA} in Figure~\ref{fig: exp_vis}.
From comparisons, three main observations can be derived:
1) removing visual cues from SoP~\cite{zhao2018the} makes it hard to do separation from the sound itself, where two recovered source spectrograms are similar among each other;
2) the quality of spectrograms generated by our method is much better than the audio-only baseline~\cite{Kong2021DecouplingMA};
3) the proposed SGN achieves competitive even better results on reconstructed spectrograms against the audio-visual approach~\cite{zhao2018the} by leveraging visual cues.
These visualizations further showcase the advantage of our simple SGN in directly learning category-aware representations from the sound itself to guide source separation.

\subsection{Experiment Analysis}

In this section, we performed ablation studies to demonstrate the benefit of introducing Source Class Tokens and Category-aware Grouping module.
We also conducted extensive experiments to explore flexible number of audio sources separation and learned sound representations.

\begin{table}[!tb]
    \renewcommand{\arraystretch}{1.1}
	\centering
	\caption{Ablation studies on Source Class Tokens (SCT) and Category-aware Grouping (CAG). }
	\label{tab: exp_ablation}
	\scalebox{0.95}{
		\begin{tabular}{ccccc}
			\toprule
			SCT & CAG & SDR & SIR & SAR \\ 	
			\midrule
			\xmark & \xmark & 1.51 & 5.92 &	6.21 \\
			\cmark & \xmark & 4.57 & 8.36 &	9.55 \\
			\xmark & \cmark & 3.32 & 7.73 &	8.56 \\
			\cmark & \cmark & \textbf{5.20} & \textbf{10.81} & \textbf{10.67} \\
			\bottomrule
	   \end{tabular}
    }
\end{table}

\noindent\textbf{Source Class Tokens \& Category-aware Grouping.}
In order to validate the effectiveness of the introduced source class tokens (SCT) and category-aware grouping (CAG), we ablate the necessity of each module and report the quantitative results in Table~\ref{tab: exp_ablation}.
We can observe that adding learnable SCT highly improves the vanilla baseline by 3.06 SDR, 2.44 SIR, and 3.34 SAR, which shows the benefit of class tokens in guiding source separation.
Meanwhile, introducing only CAG in the baseline also increases the separation results.
More importantly, incorporating SCT and CAG together into the baseline significantly raises the performance by 3.69 SDR, 4.89 SIR, and 4.46 SAR.
These results demonstrate the importance of source class tokens and category-aware grouping on extracting disentangled semantics from the audio itself for source separation.

\noindent\textbf{Generalizing to Flexible Number of Sources.}
In order to demonstrate the generalizability of the proposed SGN to flexible number of sources, 
we directly transfer the model without additional training to inference a mixture of 3 sources.
We achieve satisfactory performance of 3.18 SDR, 6.12 SIR, 5.98 SAR, and still outperform SoP~\cite{zhao2018the} (2.57 SDR, 5.56 SIR, 5.63 SAR).
These results indicate that our SGN can support to separate a flexible number of sources, which differs from existing audio-only separation methods~\cite{stoller2018waveunet,Kong2021DecouplingMA}.

\begin{figure*}[t]
\centering
\includegraphics[width=0.8\linewidth]{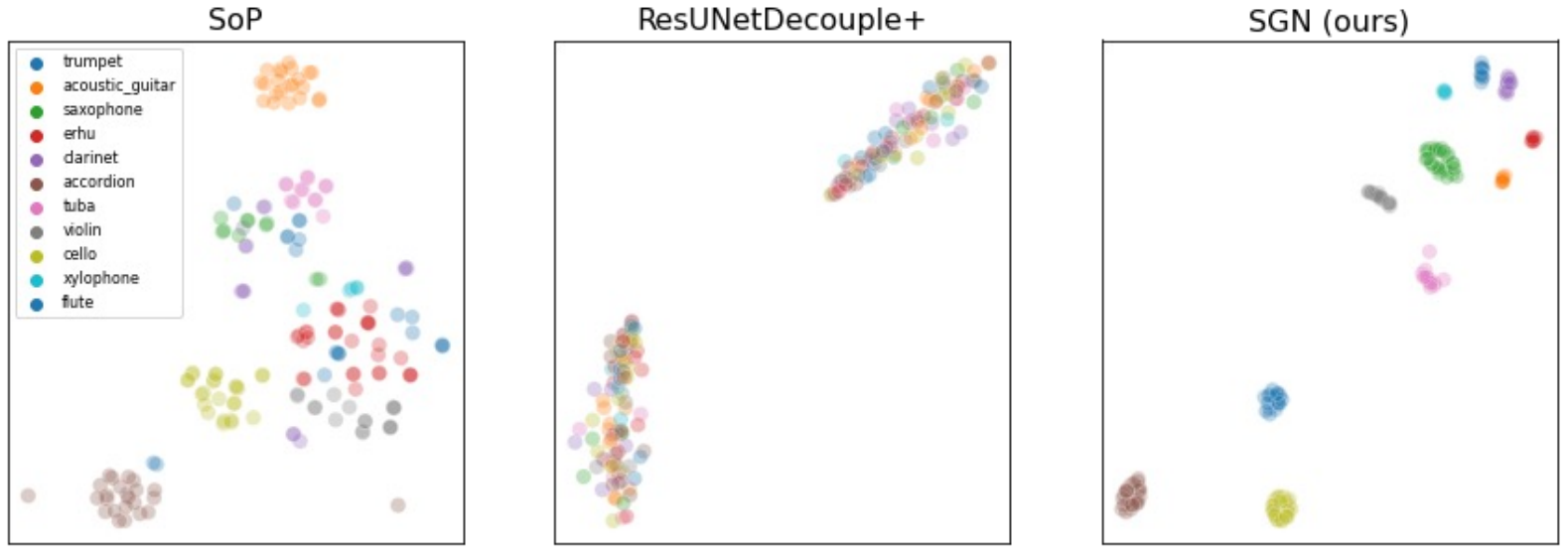}
\caption{Qualitative comparisons of representations learned by SoP, ResUNetDecouple+, and the proposed SGN. 
Note that each spot denotes the feature of one source sound, and each color refers to one category, such as ``acoustic\_guitar" in yellow and ``erhu" in red.}
\label{fig: exp_vis_feat}
\end{figure*}

\noindent\textbf{Learned Disentangled Representations.}
Learning disentangled semantic representations is essential for us to separate the sound source from a mixture.
To better evaluate the quality of learned category-aware features, we visualize the learned sound representations of 11 categories in MUSIC by t-SNE~\cite{laurens2008visualizing}, as shown in Figure~\ref{fig: exp_vis_feat}.
It is noted that each color represents one
category of source sound, such as “acoustic guitar” in yellow and “erhu” in red.
As can be seen in the last column, features extracted by the proposed SGN are both intra-class compact and inter-class separable.
In contrast to our disentangled embeddings in the semantic space, ResUNetDecouple+~\cite{Kong2021DecouplingMA} with audio as input failed to learn category-aware features, but instead learned two separate representations for magnitude and phase estimations.
With the benefit of visual cues, SoP~\cite{zhao2018the} can extract separate features on some classes, \textit{e.g.}, ``accoustic\_guitar" in yellow and ``accordion" in khaki.
However, SoP features from most classes are not separable as they did not use explicit category-aware grouping mechanism proposed in our SGN.
These meaningful visualization results further demonstrate that our SGN successfully extracts disentangled and compact representations for audio source separation.

\begin{figure*}[t]
\centering
\includegraphics[width=0.85\linewidth]{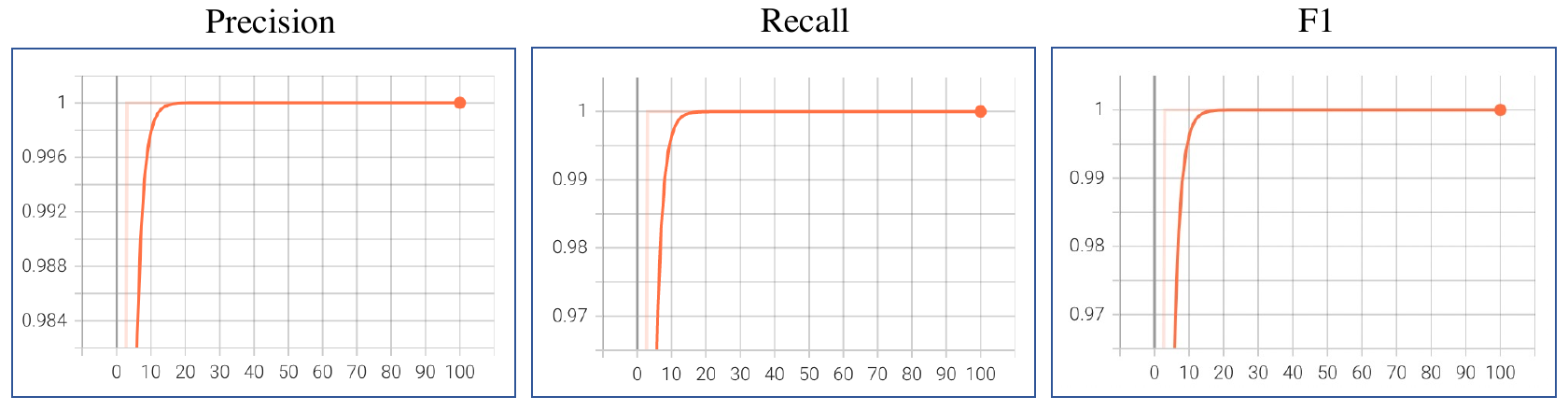}
\caption{Quantitative results (Precision, Recall, and F1 score) of learned source class tokens.}
\label{fig: vis_curve_cls}
\end{figure*}

\noindent\textbf{Quantitative Validation on Class Tokens.}
Learnable Class Tokens are essential to aggregate category-aware representations from the sound mixture.
In order to quantitatively validate the rationality of learned class token embeddings, we calculate the Precision, Recall, and F1 score of classification using these embeddings across training iterations.
We report the quantitative results in Figure~\ref{fig: vis_curve_cls}.
It can be seen that all metrics raise to 1 at epoch 20, which implies the learned class tokens have category-aware semantics.
These results also demonstrate the effectiveness of class tokens in category-aware grouping for extracting disentangled representations from audio mixtures.

\begin{table}[!tb]
    \renewcommand{\arraystretch}{1.1}
	\centering
	\caption{Exploration studies on the depth of transformer layers and grouping strategies in Category-aware Grouping (CAG) module. }
	\label{tab: exp_depth_group}
	\scalebox{0.98}{
		\begin{tabular}{ccccc}
			\toprule
			Depth & CAG & SDR & SIR & SAR \\ 	
			\midrule
			1 & Softmax & 4.32 & 8.18 & 8.07 \\
			3 & Softmax & 5.14 & 10.80 & 10.51 \\
			6 & Softmax & \textbf{5.20} & \textbf{10.81} & \textbf{10.67} \\
			12 & Softmax & 5.18 & 10.65 & 10.59 \\
			6 & Hard-Softmax & 3.25 & 6.53 & 6.36 \\
			\bottomrule
			\end{tabular}}
\end{table}

\noindent\textbf{Depth of Transformer Layers \& Grouping Strategy.}
The depth of transformer layers and grouping strategies in CAG affect the extracted and grouped representations for audio source separation.
In order to explore such effect more comprehensively, we varied the depth of transformer layers from $\{1, 3, 6, 12\}$ and ablated the grouping strategy using Hard-Softmax.
To make Hard-Softmax differentiable during training, we applied the Gumbel-Softmax~\cite{eric2017categorical,chris2017the} as the alternative.
The comparison results of the separation performance are shown in Table~\ref{tab: exp_depth_group}.
When the depth of transformer layers is 6 and using Softmax in CAG, the proposed SGN achieves the best results in terms of all metrics.
With the increase of the depth, we achieve consistently raising performance as we extract better audio representations from the mixture.
However, increasing the depth to 12 will not continually improve the result since 6 transformer layers might be enough to extract the learned embedding for category-aware grouping.
In addition, replacing Softmax with Hard-Softmax significantly deteriorates the separation performance of our approach, which implies the importance of the proposed CAG in extracting disentangled semantics for separation.

\begin{figure*}[t]
\centering
\includegraphics[width=0.98\linewidth]{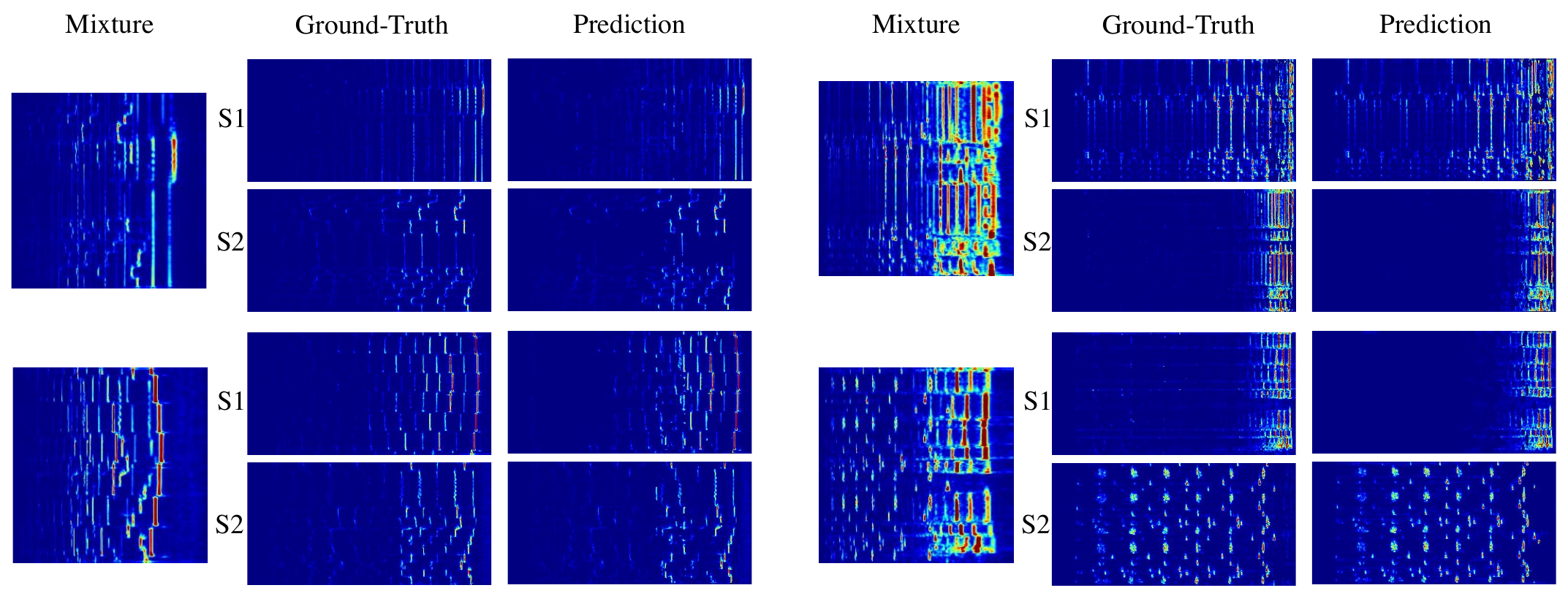}
\caption{Visualization results of separated source spectrograms.}
\label{fig: vis_sep}
\end{figure*}

\noindent\textbf{More Qualitative Visualization on Separation.}
In order to qualitatively demonstrate the effectiveness of our method, we report more visualization results in Figure~\ref{fig: vis_sep}.
We can observe that the proposed SGN achieves decent separation performance in terms of reconstructing the spectrogram for each sound source.

\subsection{Limitation}

Although the proposed SGN achieves superior results on music and universal sound source separation, the gains of our approach over audio-visual models are not significant.
One possible solution is to incorporate the visual modality with audio features together for multi-modal grouping.
Meanwhile, we notice that if we transfer our model directly without additional training, it would be hard to separate all unseen classes as we do not have learned unseen class tokens to guide source separation.
The primary cause is that we need to pre-define a set of category types for training.
Therefore, the future work is potentially to learn enough large number of sound source tokens or to explore continual learning when it comes to new classes of audio.

\subsection{Different from GroupViT and our SGN}

When compared to GroupViT~\cite{xu2022groupvit} on image segmentation, there are three main distinct characteristics of our SGN for addressing audio source separation, which are highlighted as follows:

\noindent 1) \textbf{Disentangled Constraint on Class Tokens}. 
The most significant difference is that we have learned disentangled class tokens for each sound source, \textit{e.g.}, 11 tokens for 11 categories in MUSIC dataset.
With the disentangled constraint, each class token does not have overlapping information to learn during training, where we apply a cross-entropy loss
$\sum_{i=1}^C\mbox{CE}(\mathbf{h}_i, \mathbf{e}_i)$, with an one-hot encoding target $\mathbf{h}_i$ to constraint each category probability $\mathbf{e}_i$.
In contrast, the number of group tokens used in GroupViT is indeed a hyper-parameter and they do not apply any constraint on them.

\noindent 2) \textbf{Category-aware Grouping}. 
We introduce the category-aware grouping module for extracting individual semantics from audio spectrogram in the original mixed space.
However, GroupViT utilized the grouping mechanism on visual patches without class-aware tokens involved.
Thus, GroupViT can not be directly applied on sound spectrogram for solving source separation.
In addition, they leveraged multiple grouping stages during training and the number of grouping stage is a hyper-parameter.
In our case, only one category-aware stage with meaningful class tokens is enough to learn distinguished representations in the semantic space.

\noindent 3) \textbf{Class as Weak Supervision}.
We apply the source class as weak supervision to address sound separation problem, but GroupViT used a contrastive loss to match the global visual representations with text embeddings.
Therefore, GroupViT required a large batch size for self-supervised training on large-scale data.
In this work, we do not need the unsupervised learning on large-scale simulated sound data with expensive training costs.

\section{Conclusion}

In this work, we propose SGN, a novel Semantic Grouping Network that directly disentangles audio representations and extracts high-level semantic information for each sound source from input mixtures.
We introduce learnable class tokens of sounds to aggregate category-wise source features. 
Then, we use the aggregated semantic features as the guidance to separate the corresponding audio sources from the mixture. 
Different from previous sound separation approaches, the proposed SGN can support separation from a flexible number of sound sources and is capable of generalizing to tackle sources from different audio domains. 
Empirical experiments on both music-only and universal sound separation benchmarks demonstrate the significant advantage of our SGN against audio-only methods and audio-visual models without additional visual cues involved.

\noindent\textbf{Broader Impact.}
The proposed approach separates the mixture sounds from manually-collected music and universal datasets, which might cause the model to learn internal biases in the data.
For instance, the model could fail to separate rare but crucial sound sources.
Therefore, these issues should be addressed for the deployment of real applications.

%

\bibliography{reference}
\bibliographystyle{IEEEtran}

\vfill

\end{document}